# ON THE PREDICTION OF THE OCCURRENCE DATES OF GLEs

Jorge Pérez-Peraza, Alan  Juárez-Zúñiga,  Julián Zapotitla-Román
Instituto de Geofísica, Universidad Nacional Autónoma de México, C.U., Coyoacán, 04510, México, D.F. MEXICO

Manuel Alvarez-Madrigal
Instituto Tecnológico de Monterrey, Campus Ciudad de México, Calle del Puente 222, 14380, Mexico D.F.. MEXICO

**Ground level enhancements (GLEs) are relativistic sol particles measured at ground level by a worldwide network of cosmic ray detectors. These sporadic events are associated with solar flares and are assumed to be of a quasi-random nature. Their study gives us information about their source and propagation processes, about the maximum capacity of the sun as a particle accelerator engine, about the magnetic structure of the medium traversed, etc. Space vehicles may be damaged by this kind of radiation, as well as electric transformers and gas pipes at high latitudes. As a result, their prediction has turned out to be very important, but because of their random occurrence, up to now few efforts to this end have been made. The results of these efforts have been limited to possible warnings in real time, just before GLE occurrence, but no specific dates have been predicted well enough in advance to prevent possible hazards. In this study we show that, in spite of the quasi-stochastic nature of GLEs, it is possible to predict them with relative precision, even for future solar cycles. We reproduce previous GLE events and present results for future events.**

## INTRODUCTION

GLEs of relativistic solar protons (RSPs) are sporadic phenomena that, to a certain extent,

follow the time behavior of the 11-year cycle of solar activity (SA); however, they do not follow the intensity of the SA cycle: for instance, cycle 23 had more GLE events than cycle 22, which was a much more intense one. In total, 71 GLEs have been recorded: the first measurement was on 28 February, 1942 (GLE01) and the last one, on 17 May, 2012 (GLE71). Though the average occurrence rate is ~ 0.99 year$^{-1}$, the span between events may sometimes be almost 6 years, as was the case between GLE70 and GLE71.

The sequence of Magnetohydrodinamic processes that takes place in the subphotosphere and the other solar atmospheric layers demonstrates a very complex evolution in time and space. A huge amount of effort has been expended for many decades to explain this evolution. However, up to the present, only partial aspects of it can be understood, and of course very few prognoses can be made with these theoretical models in order to predict when a solar flare producing relativistic particles will occur. It is often assumed that GLEs are random phenomena. As a result, the effort to predict GLE's has been limited to attempts based on a real-time survey[1], requiring an organized system of neutron monitor detectors, coupled to computers having specific algorithms, which, in the best of cases, would only provide information minutes or hours before the GLE occurrence.

By means of the analysis of GLE data series, we have shown[1,2], however, that GLEs maintain a cyclic tendency represented by harmonic signals and have determined GLE intrinsic periodicities: mid-term periodicities (on the order of months and years), short-term periodicities (on the order of days), and ultra-short periodicities (in the order of minutes and hours). A wavelet-coherence analysis between the GLE series and the photospheric as well as coronal series indicates that most of the periodicities mentioned above are present from the sub-photospheric to the coronal layers. Such synchronization seems to indicate that GLE production is not an isolated local phenomenon but involves global regions of the sun's atmosphere. This fact seems to argue against the full-stochasticity of GLEs, however, up to now no consistent theory can prove it.

On the other hand, intrinsic harmonics of Galactic Cosmic Rays (GCR) seems to act as precursors of GLE's, so that, by ignoring the complex physics involved and using only the GCR periodicities, we have developed here a semi-empirical method for the prediction of the appearance of GLEs several months, even years, in advance.

## DATA AND ANALYSIS

Data on the GLEs and GCR are furnished by the worldwide network of neutron monitor (NM) and muon telescope (MT) stations. Data from 1942-1964 are limited to hourly and daily values and come from a reduced number of stations. For the goal of this research a resolution of daily values is quite enough. Data since 1964 with high reliability are available with much higher resolution from many NM stations; for this specific period we have used data from the Oulu station.

To determine the main oscillation periodicities for non-stationary series, such as those for GCR, as well as their time evolution, we apply here the Morlet wavelet technique[3]. This is a very well-known tool for analyzing localized variations of power within a given time series at many different periodicities, when one is dealing with a non-stationary series and the coherence between two non-stationary series. The statistical significance level is estimated using Monte Carlo methods with red noise. The so-called global wavelet spectrum (GWS) is an average of the power spectra at each resolution level, i.e., it assumes that the time series has an average power spectrum relative to the red noise of Fourier: harmonics above this average spectrum (the slashed line) in Fig. 1 represent real signals with levels of reliability higher than 95%. The importance of the GWS is the distribution of signals with the same characteristics to determine which harmonics contain greater power[4].

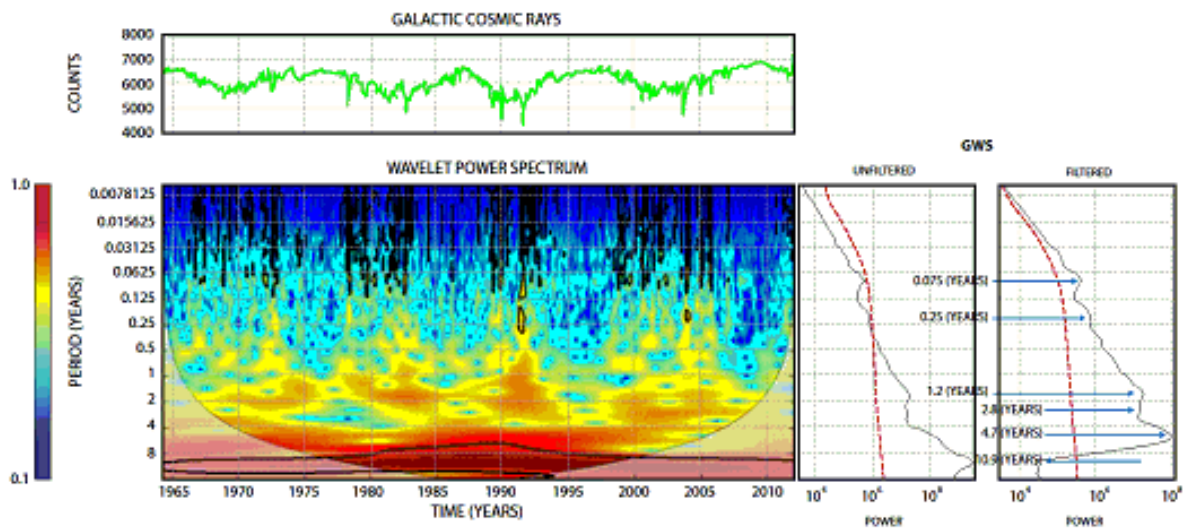

Figure 1.  **Spectral Analysis:** Upper panel is  the GCR flux, middle panel is the wavelet spectrum and right panels is the global energy spectrum before and after filtering.

We apply wavelet analysis to the series of GCR daily data (top of Fig. 1), obtaining their wavelet spectrum and global-energy spectrum. From these we obtain the most confident periodicities, which are in the range of 0.075 to 16 years, where the most prominent one, i.e. the controlling pulse from the energetic point of view, is at 10.9 years, as indicated by the GWS in Fig. 1.  We find that the 10.9 year periodicity allows for a classification of the 70 events into 6 groups, (plus an incipient one, group 7) as is shown in Fig. 2. The first group is somewhat uncertain, because we do not know if the event of February 28, 1942 was the first of r group 1.

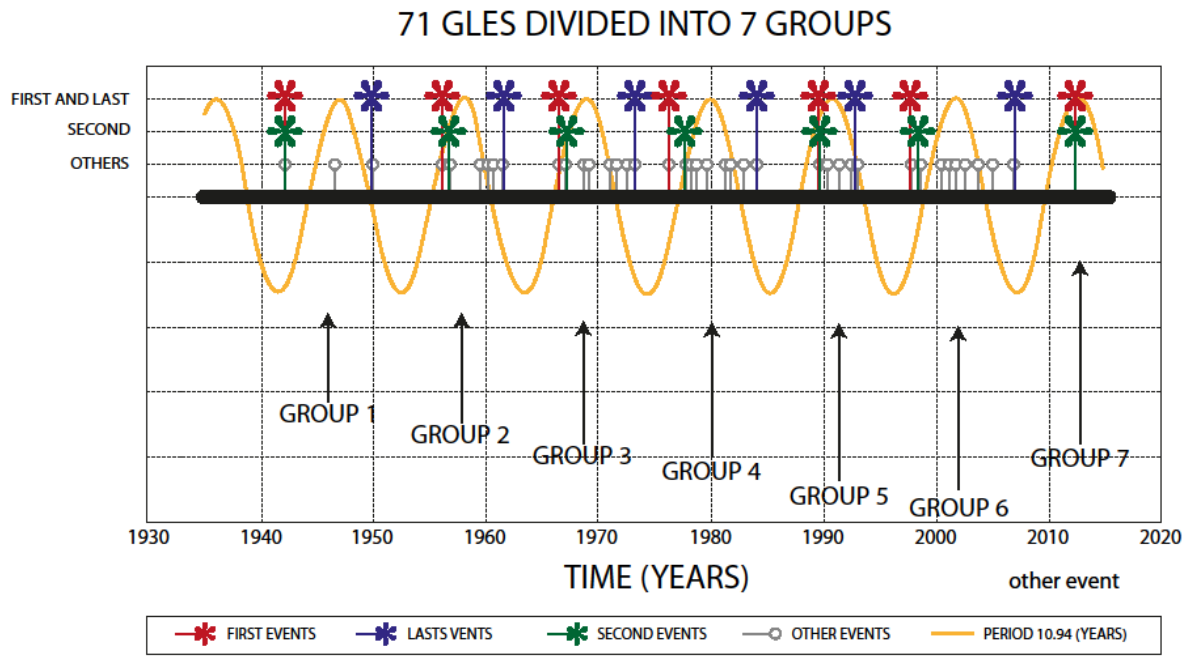

Figure 2. **Classification**: Grouping of GLE into six groups (plus the incipient group 7) ccording to their predominant harmonic at 10.94 years.

In order to discern high frequencies, we apply the Daubechies filter[5] to remove the 10.94 harmonic, which has quite a high energy content and thus hides shorter periods. It should be mentioned that at the interior of a given group of GLEs events may occur sometimes very close among them, which is the reason for the need of a fine structure of the GCR harmonics. Once the dominant harmonics have been identified, they are applied to signals from the discrete Fourier transform obtaining the amplitude and phase spectra of the different periodicities.

The amplitude is normalized in the timeline because of the irrelevance of other properties such as intensity, profile, particle energy, stabilization, etc. We look for their differences in states only on the date of occurrence.

From the obtained harmonics, we select those that meet the criteria[2] imposed a priori: for a selected harmonic, the occurrence date of a given kind of event (for instance, the first, second, or last) must fall in the positive part of the harmonic in all the groups, or, alternatively, in the negative part of the selected harmonic in all the groups. Under these conditions, a new function is created for each harmonic by applying the unitary step selection. The obtained functions for the selected harmonics in a given kind of GLE are multiplied to obtain their intersection and, in this way, to determine the possible regions of occurrence of a given kind of GLE. For instance, for the first GLE in Group 7, the region of probability is found in regions delimited by the positive part of the harmonic associated with the 1.2 and the 10.94 year periodicities for all the groups, and by the negative part of the harmonic in the case of the 4.7 year periodicity for all the groups (see Fig. 4). Because the harmonics are chosen by behavioral observations of the occurrences of the given kind of GLE, in Fig. 3 we illustrate the reconstruction of the occurrence time intervals for the first, second and last GLEs of all groups. It can be seen in Fig. 3 that the predicted regions correspond to intervals where events occurred in groups of 1 to 6 (interpolation ). The red and blue lines in Fig. 3 indicate the exact dates of GLE occurrence. The dashed areas indicate the predicted date intervals. The fact that it is precisely in Group 1 that the intervals are narrower seems to indicate that the 28 February event was effectively the first event of that group. It must be noted that for incipient Group 7 and future Group 8, there are several possibilities for time interval occurrence; a way to select the optimum one will be discussed in the next section in terms of statistical analysis.

Since we assume that the behavior is harmonic, we are able to find the occurrence intervals for future GLEs (through extrapolation) and even of groups before Group 1, when the neutron monitors were not yet in operation. It is in this way that we predicted[2] the occurrence date of GLE71 (the first one in Group 7); it can be seen in Fig. 4 and in Table 1 that this method leads to two different time intervals, the first of which is out of the question, because before 2012, GLE71 had not appeared.

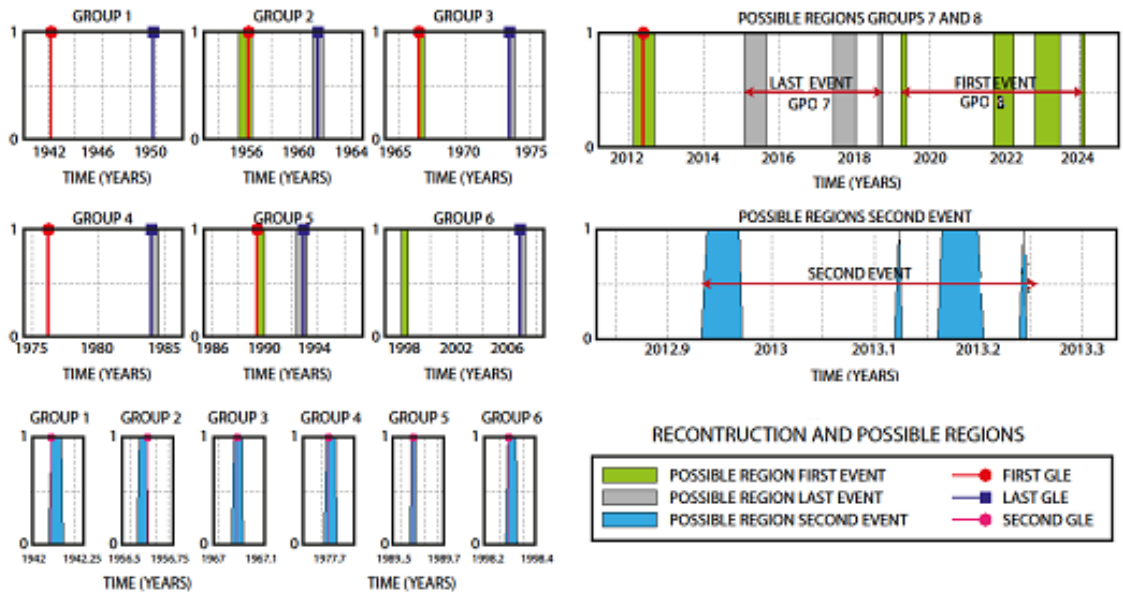

Figure 3. **Interpolation and Extrapolation;** Reconstruction of occurrence dates of first, second and last GLEs for the past 6 groups, the current one and the next one.

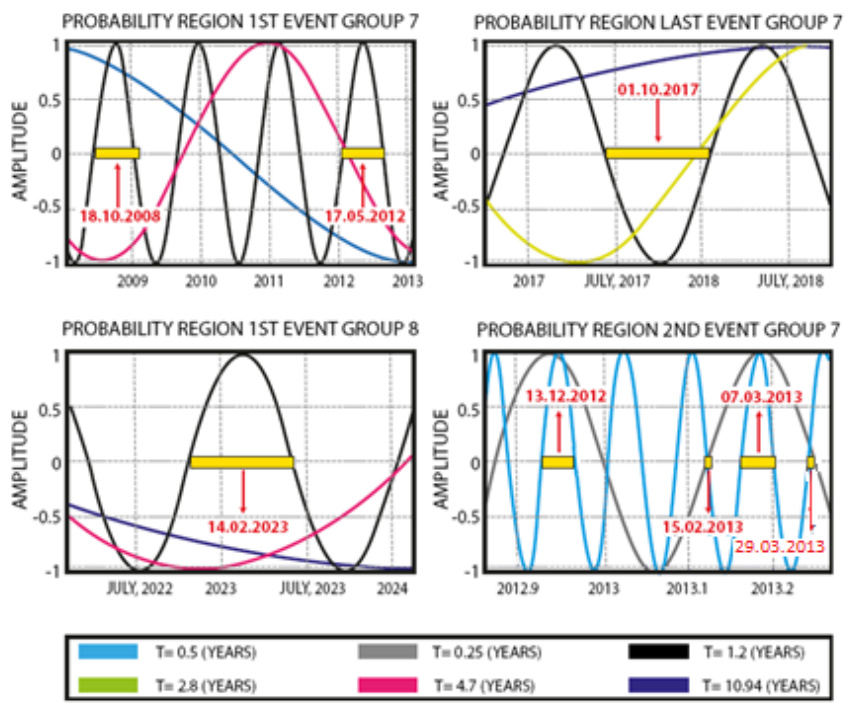

Figure 4. **Prognosis**: Predicted time intervals for occurrence of GLEs for Group 7 and future Group 8.

For the second GLE of Group 7 there are three possibilities, one of which is out of the question, since it falls during 2012. For the last GLE of Group 7 and the first one of Group 8, there are several possibilities, for which statistical weights must be assigned, as described in the next section. For the second and last GLEs of Group 7 and the first GLE of Group 8, Fig. 4 illustrates the predicted occurrence time intervals, but only for those with the highest statistical weights.

**STATISTICAL WEIGHTS OF THE PREDICTED DATES OF GLE OCCURRENCET**

The methodology we developed for predicting time intervals of GLEs occurrence gives different possible time intervals for the occurrence of a GLE. We have implemented a procedure to select the interval of highest probability based on the available data. The procedure was applied as described below:

With the occurrence dates of the GLEs we obtain the intervals of potential occurrence of the studied events. For the first event of a new group (Table 1 and Table 2), the main indicator is the elapsed time between the first event of each group and last event of the previous one, so for the first events of groups 7 and 8 we considered the average lag between the last and first events among consecutive groups This average is designated hereafter as (FIRST - LAST). On the other hand, since the dominant periodicity that controls the first and last events of each group is that of 10.94 years, we used the average time elapsed between first events among consecutive groups namely (FIRST-FIRST).

| PROBABILITY OF POSIBLE INTERVALS FOR THE 1st EVENT OF THE GROUP 7 | | | | |
|---|---|---|---|---|
| INTERVAL | LIMITS | | FIRST - LAST PROBABILITY | FIRST - FIRST PROBABILITY |
| 1 | START | 28.09.2007 | 0.01327918 | 19.5570768 |
|   | END | 26.11.2007 |   |   |
| 2 | START | 02.07.2008 | 0.08935983 | 8.79632197 |
|   | END | 04.02.2009 |   |   |
| 3 | START | 06.03.2012 | 99.8973487 | 71.6478161 |
|   | END | 04.09.2012 |   |   |
| TOTAL |   |   | 100 | 100 |

**Table 1.** Predicted time interval for the first GLE of Group 7 (GLE71): 06.03.2012 – 04.09.2012, that has taken place on 17.05.2012

| PROBABILITY OF POSIBLE INTERVALS FOR THE 1st EVENT OF THE GROUP 8 | | | | |
|---|---|---|---|---|
| INTERVAL | LIMITS | | FIRST - LAST PROBABILITY | FIRST - FIRST PROBABILITY |
| 1 | START | 31.03.2019 | 0.00020838 | 0.00544496 |
|   | END | 04.05.2019 |   |   |
| 2 | START | 23.09.2021 | 17.2140844 | 10.5093638 |
|   | END | 24.03.2022 |   |   |
| 3 | START | 29.10.2022 | 73.0666246 | 63.8089932 |
|   | END | 03.06.2023 |   |   |
| 4 | START | 08.01.2024 | 9.71908263 | 25.6761981 |
|   | END | 11.02.2024 |   |   |
| TOTAL |   |   | 100 | 100 |

**Table 2.** Predicted time interval for the first GLE of Group 8: 29.10.2022 – 03.06.2023.

e groups, called (LAST- LAST).

| PROBABILITY OF POSIBLE INTERVALS FOR THE LAST EVENT OF THE GROUP 7 LAST-LAST | | | |
|---|---|---|---|
| INTERVAL | LIMITS | | PROBABILITY |
| 1 | START | 25.01.2015 | 9.50870897 |
|   | END | 30.08.2015 |   |
| 2 | START | 15.06.2017 | 87.829888 |
|   | END | 18.01.2018 |   |
| 3 | START | 25.08.2018 | 2.66140301 |
|   | END | 02.09.2018 |   |
| TOTAL |   |   | 100 |

**Table 3.** Predicted time interval for the last GLE of Group 7: 15.06.2017 – 18.01.2018.
first one is already over.

For the final event of group 7 (Table 3), we used the average time between the last events in consecutive groups, called (LAST- LAST).

For the second event in Group 7 (Table 4), we used the average time interval between the second and first events, (SECOND-FIRST), from the six previous groups. In this case we show three possible intervals, from which we assume that the second is the most probable, because the first one is already over.

| INTERVAL | LIMITS | | PROBABILITY |
|---|---|---|---|
| colspan="3" | PROBABILITY OF POSSIBLE INTERVALS FOR THE 2nd EVENT OF THE GROUP 7 (SECOND-FIRST) | | |
| 1 | START | 15.02.2013 | |
|  | END | 17.02.2013 | 6.70118607 |
| 2 | START | 02.03.2013 | |
|  | END | 16.03.2013 | 87.5095287 |
| 3 | START | 29.03.2013 | |
|  | END | 30.03.2013 | 5.78928522 |
| TOTAL |  |  | 100 |

**Table 4.** Predicted time interval for the second GLE of Group 7: 02-03.2013-16.03.2013.

In each of the previously mentioned cases we obtained their standard deviations of the elapsed times. The lower the standard deviation the higher the accuracy in calculating the probability. Using the average and the standard deviations the probability function was calculated within each interval. The normal probability density function indicates what the interval with the highest accumulated area under the function will be. The sum of the cumulative probabilities of all intervals must be 100%.

Because of the small size of the data set, to calculate a statistical weight to each possible interval of occurrence of a GLE, we assume that each evaluated parameter (differences between the dates of events of interest, last-first, first-first, last-last of consecutive groups and first-second in the same group) follows a normal probabilistic function, according to the central limit theorem, whose values for the mean and standard deviation can be obtained from the samples, even if the number of these is small. For each data set there are two confidence intervals of 95%, one for the mean and other for the variance, such that we can affirm that the true values of these parameters are within these ranges. In order to select the better value more accurately, we proceeded to evaluate the certainty of each combination of media and variance within the confidence intervals that describe the known events. This was determined as follows:

Confidence intervals obtained for the mean and the variance are continuous; for the evaluation of each value of media and variance we need the evaluation of point values belonging to each interval, we therefore discrete confidence intervals. We arbitrarily select in ten equidistant values within the interval in order to assess their goodness to predict the events.

We select each pair of mean and variance and compute with them the normal distribution of probability of occurrence of the different possible intervals for a given event in each of the 6 groups. After matching the observed higher statistical weight range with known event occurrence, ie, the probabilities of success of the prediction by counting the number of events that have occurred within the most likely interval. To express it in terms of probabilistic numeric values , we divide the number of success by the total of known events (6 or 7 reference events depending on the event type): 6 for the case of the first event of group 7; seven for the last event of group seven and 7 for the first event of group eight (assuming we know the first and last events of the group seven). Finally we assign to the evaluated pair mean-variance the value of the probability of success.

Doing this to all pairs of mean- variance the Sensitivity maps can be constructed (Figure 5-9). In these contour plots the best values of mean and the variance that are more useful in terms of accuracy for predicting future events can be determined. Thus, we select the best values of the parameters of the normal distribution to determine the statistical weight of the predictions. The sensibility of hit in the prediction can be analyzed in terms of different values of the media and the variance: as it is expected, It can be seen from the maps, that the higher number of data, the smaller the region containing the assumed real values of the mean and variance, and so, the higher the probability of success.

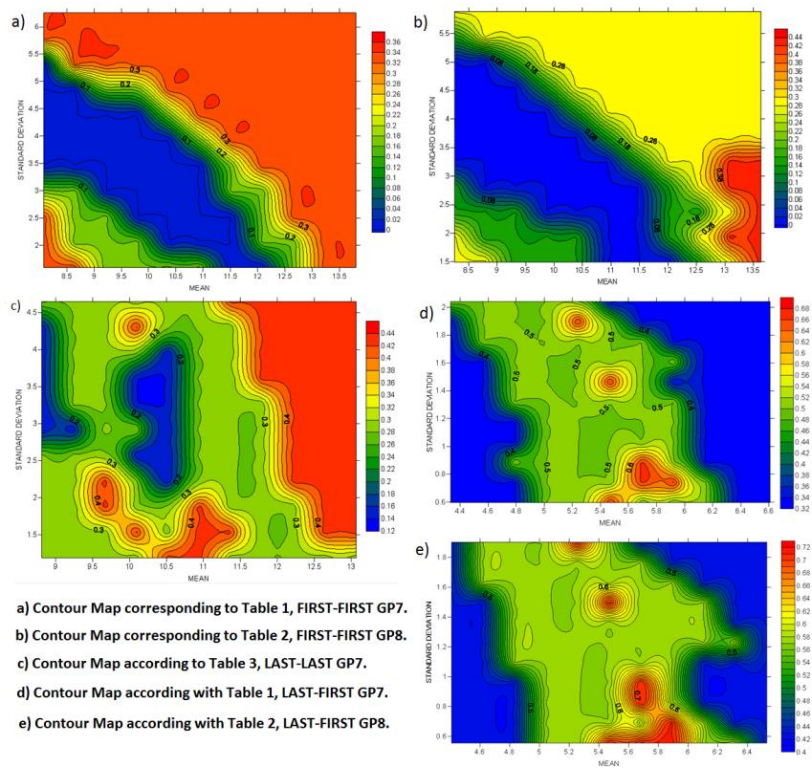

a) Contour Map corresponding to Table 1, FIRST-FIRST GP7.
b) Contour Map corresponding to Table 2, FIRST-FIRST GP8.
c) Contour Map according to Table 3, LAST-LAST GP7.
d) Contour Map according with Table 1, LAST-FIRST GP7.
e) Contour Map according with Table 2, LAST-FIRST GP8.

Fig. 5 . Sensibility Map of Probabilities

For the second event of group 7, we assume that the mean and variance of the sample are suitable for calculation of probability; due to the fact that in this case there are many possible areas of occurrence a sensitivity analysis is not useful because it provides no information on the certainty of the prediction as in the previous cases, Therefore, we assume that only the regions predicted within the interval $\bar{x} \pm \sigma$ are likely to occur. These are the regions that are after the first event group and up to about a year later. This probability is assigned by the normal function of cumulative probability. Within these ranges it is noted that the probability distribution tends to be uniform. We assume that the probability of occurrence of an interval is independent from the others, so by the time an interval is past, its probability will be zero and the probability of occurrence of the event is distributed among the remaining intervals, such that their sum will be always the expected 100% (i.e. normalized to 1).

**DISCUSSION**

It has been found that the first and last GLEs of each group are mainly controlled by the 10.94 year harmonic, such that all groups are connected among them by this harmonic. On the other hand, it is observed that at the interior of a given group, there always occurs several GLE's from the beginning of that cycle up to the maximum of the 10.94 year cycle, and such interior GLEs are closely related to the harmonics of one to three months. There is a lag between groups ( ~ 5.5 years) roughly concurring with the maxima of solar activity, but once a group is initiated by a given *trigger mechanism* (probably similar for every group) then, it seems that another *discharger process* begins to *shoot* GLEs up to the end of the 10.94 years cycle, followed again by a lag of ~ 5.5 years and so on.

The presented method for long term prediction of GLEs has limited scope; though mathematically we could predict events through thousands of years in the past and the future, the obtained dates would be entirely fictitious, because as one moves away of the observation dates a significant error is accumulated that keeps us away from reality. The higher confidence is for a couple of decades before and after the observation dates. Besides, the greater accuracy of the method is for the prediction of first and last events, controlled by the same harmonic; since not all groups have the same number of events the configuration of harmonics describing events at the interior of a given group is not necessarily the same from group to group. An additional restriction concerning intermediate events is that these must be separated from each other by at least twice the minimum observation frequency

(frequency Nyquits), otherwise, we would be violating a fundamental theorem of Signal Analysis. In order to test the accuracy of our predictions, an analysis of probabilities was performed by using the normal probability distribution, whose parameters, mean and standard deviation were found by confidence interval estimation of the sample measures; this was done with t-student and chi-square distributions respectively. We select those that provide the highest probability of success.

## SUMMARY

In the absence of a physical theory to predict with satisfactory accuracy the production of relativistic particles on the sun, with effects at ground level (GLEs), several alert systems in real time have been proposed but have not yet been implemented. Such alert systems would provide information for the minutes or hours previous to the GLE ***occurrence. In order to overcome such a fallacy, we propose a method here based on spectral analysis and statistical analysis to predict, the occurrence dates of proximate and future GLEs, as well as those that occurred before the advent of cosmic ray detectors. GLEs have been classified in groups, each one with a duration of 10.94 years. Using the proposed method, a reconstruction of the 70 past GLEs and the prediction of first one of the present group has been done with a high accuracy. In spite that the prediction of internal GLEs within a given group is not so accurate, as it was for the first event of group-7, when we know the exact date of GLE's of the previous group., we also predict here (with a lower probability of only 48.8%) the occurrence of the next GLE, the second one in Group 7, to occur during the month of March 2013. For the last event of this group, we predict its occurrence between 15 June, 2017 and 18 January 2018. For the first GLE of the next group (Group 8), the prediction is for the interval between 29 November, 2022 and Jun 3, 2023. It should be noted that these time intervals are not deterministic but compete with others of lower probability that from a statistical point of view cannot be disregarded.